\documentclass[11pt]{article}
\usepackage{amsmath}
\usepackage{amsfonts}
\usepackage{amssymb}
\usepackage{mathptmx}
\usepackage{slashed}
\usepackage{latexsym}
\usepackage{graphicx}
\usepackage{color}
\usepackage{float}
\usepackage{multirow}

\newcommand{\newc}{\newcommand}
\newc{\beg}{\begin}
\newc{\beq}{\beg{equation}}
\newc{\eeq}{\end{equation}}
\newc{\bea}{\beg{eqnarray}}
\newc{\eea}{\end{eqnarray}}
\newc{\m}{\mathcal}
\newc{\nn}{\nonumber}
\newc{\diag}[1]{\ensuremath{{\rm diag} \left( #1 \right)}}

\begin{document}

\begin{titlepage}

\vspace*{0.7cm}

\begin{center}
{
\bf\LARGE
Non-universal $Z'$ from Fluxed GUTs}
\\[12mm]
Miguel Crispim Romao$^{\star}$
\footnote{E-mail: \texttt{mcr1n17@soton.ac.uk}},
Stephen~F.~King$^{\star}$
\footnote{E-mail: \texttt{king@soton.ac.uk}},
George~K.~Leontaris$^{\dagger}$
\footnote{E-mail: \texttt{leonta@uoi.gr}},
\\[-2mm]

\end{center}
\vspace*{0.50cm}
\centerline{$^{\star}$ \it
School of Physics and Astronomy, University of Southampton,}
\centerline{\it
SO17 1BJ Southampton, United Kingdom }
\vspace*{0.2cm}
\centerline{$^{\dagger}$ \it
Physics Department, Theory Division, Ioannina University,}
\centerline{\it
GR-45110 Ioannina, Greece}
\vspace*{1.20cm}

\begin{abstract}
\noindent
We make a first systematic study of non-universal TeV scale neutral gauge bosons $Z'$ arising naturally from a class of F-theory inspired 
models broken via $SU(5)$ by flux. 
The phenomenological models we consider may originate from semi-local F-theory GUTs arising from a single $E_8$ point of local enhancement, assuming 
the minimal ${\cal Z}_2$ monodromy in order to allow for a renormalisable top quark Yukawa coupling. We classify such  
non-universal anomaly-free $U(1)'$ models requiring a minimal low energy spectrum and also allowing for a vector-like  family. 
We discuss to what extent such models can account for the anomalous $B$-decay ratios $R_{K}$ and $R_{K^*}$.
 \end{abstract}

 \end{titlepage}

\thispagestyle{empty}
\vfill
\newpage

\setcounter{page}{1}

\section{Introduction}

In the Standard Model (SM) 
of strong and electroweak interactions the gauge couplings to lepton fields  are flavour independent. This is a well 
known property of the SM gauge interactions which is usually called Lepton Flavour Universality (LFU).  Among the various   
processes that can be used to  test LFU, are the flavour changing decays involving quark fields such as $b\to s\ell^+\ell^-$ 
where $\ell^{\pm}$ stand  for $e^{\pm}, \mu^{\pm}, \tau^{\pm}$ leptons. Recent experimental results~\cite{Aaij:2017vbb} 
indicate violations  of LFU which may be due to a new neutral boson $Z'$~\cite{Langacker:2008yv,Cvetic:2011iq}  which both changes quark 
flavour, namely $b\rightarrow s$, and couples predominantly to muons rather than electrons, with left-handed quark and lepton 
couplings preferred~\cite{Hiller:2017bzc}. The flavourful $Z'$ models are many and varied. One attractive possibility is to 
consider a vector-like family with non-universal $Z'$ families which mixes  with the three chiral families of quarks and leptons 
with universal $Z'$ couplings, thereby introducing non-universality  by the back door~\cite{King:2017anf}~\footnote{Extensive 
 references to other non-universal $Z'$ models are also included in~\cite{King:2017anf}.}.

It is commonly believed that the SM is not the final theory of fundamental interactions, but just an effective low energy 
limit of some (partially) grand unified theory (GUT), possibly embedded in a string scenario, leading to new physics phenomena 
and deviations from the SM predictions. For example, new gauge bosons $Z'$ associated with additional abelian  symmetries 
 arise from such string embeddings~\cite{Langacker:2008yv}.  Within the framework of F-theory~\cite{Vafa:1996xn} there are 
 many such extra gauged abelian symmetries~\cite{Beasley:2008kw}, which however are normally assumed to be broken at the high 
 energy scale. Usually these gauge bosons are assumed to have universal couplings to quarks and leptons~\cite{Karozas:2017hog},
however there is no reason in principle from the point of view of F-theory why this should be the case. Despite this there has been 
no systematic study of non-universal $Z'$ gauge bosons at the low energy scale arising from F-theory to our knowledge.

Motivated by the above recent hints for non-universality, we make a first systematic study of anomaly-free non-universal $Z'$ models arising from F-theory.
From a phenomenological point of view, we refer to such models as fluxed GUTs. These models assume a set of rules extracted from F-theory 
and which can form the basis for a phenomenological analysis of models. In particular we shall use these rules to search for a set of anomaly 
free models~\footnote{For recent anomaly free models with non-universal $Z'$- couplings see~\cite{Bonilla:2017lsq,Ellis:2017nrp}.}
in which the low energy effective theory consists of the Minimal Supersymmetric Standard Model (MSSM) augmented by an extra 
gauged  $U(1)'$ group broken at the TeV scale, together allowing also for a possible vector-like family at that scale. 
We shall discover that none of the models with a minimal low energy matter spectrum is capable of explaining the 
recent B-physics anomalies, while it is possible to account for these anomalies by allowing
an extra vector-like family as in \cite{King:2017anf}.

The layout of the paper is as follows. In section two we introcude the notion of fluxed GUTs which are based on essential features  playing instrumental r\^ole in local F-theory constructions.  In particular, we describe 
the r\^ole of fluxes on the breaking of the gauge symmetry, the splitting of GUT representations and the fermion chirality.  
In section three we implement these rules to construct  $SU(5)$ GUTs
with additional abelian symmetries allowing non-universal couplings to fermions. We focus on  specific  models  where both $SU(5)$ GUT and abelian symmetries are 
embedded in $E_8$ and formulate the anomaly cancellation conditions which constrain the  spectrum of the emerging effective models. In section four we  present the solutions of these conditions and classify them according to
the properties of the resulting light particle spectrum. We first present characteristic  examples of $SU(5)$ with  the minimal supersymmetric spectrum 
having non-universal couplings with the extra $U(1)$.  We also show solutions
containing vector-like fermions with non-universal couplings and discuss 
their implications to low energy physics. In section five we present our conclusions.

\section{Fluxed GUTs}

In our quest for a possible interpretation of such kind of experimental results with the incorporation of a new gauge 
boson  $Z'$,  we will examine  GUT models that can in principle be derived in the context of F-theory model building~\cite{Beasley:2008kw}. 
We stress that this  particular  string theory framework  provides the appropriate  tools for a natural inclusion
 of abelian gauge bosons with non-universal gauge couplings to fermions. Indeed, among other basic ingredients of
 F-theory constructions, a prominent r\^ole for the incarnation of the aforementioned scenario is played by the 
 particular structure of the compact manifold and the (abelian) fluxes.  
 
 \noindent
 We recall first that the compactification space  consists of an elliptically fibred Calabi-Yau manifold
($CY_4$ fourfold)  of four  complex dimensions  whose geometric singularities display a group structure and,  in F-theory, 
  are associated with the gauge symmetries of the effective field theory model~\cite{Vafa:1996xn}.  We will assume that 
  the internal manifold has a structure characterised by the maximal exceptional symmetry $E_8$,  where 7-branes
  wrap an appropriate divisor accommodating the $SU(5)$  gauge group of the effective field theory.  
  In this context, the zero mode spectrum  descends from the decomposition of the $E_8$ adjoint, therefore the 
  following picture emerges in   the effective model
 \bea
 E_8 & \supset & SU(5)_{GUT} \times SU(5)_{\perp} \label{SU(5)}\label{E8dec} \\
 248 & \rightarrow & (24,1)+(1,24) +\boxed{ (10,5)+(\overline{5},10)}+(\overline{5}, \overline{10})+(5,\overline{10})
 \label{E8adj}\,,
 \label{DP00}
 \eea
 where in~(\ref{E8dec}), the group factor $SU(5)_{GUT}$, is identified with the grand unified symmetry 
 of the effective model, and in~(\ref{E8adj}) the representations in the box contain the $SU(5)_{GUT}$ fields.
 As is well known, ordinary matter is accommodated in 
 $10$ and $\overline{5}$  representations which, in the present construction, appear with non-trivial
 transformation properties under the (perpendicular) symmetry $SU(5)_{\perp}$.  For the purposes of
 the effective field theory description, it is adequate to work in the Higgs bundle picture and 
 express the transformation properties of the $SU(5)_{GUT}$ matter in terms of the Cartan generators
 of $SU(5)_{\perp}$ and the five weights $t_i, i=1,2,\dots, 5$ satisfying the $SU(5)$ tracelessness condition
 \beq
 t_1+t_2+t_3+t_4+t_5=0~\cdot 
 \eeq
 In this case, the  superpotential can be maximally constrained by four $U(1)_{\perp}$'s  (``perpendicular'' to $SU(5)_{GUT}$) 
 according to the breaking pattern
 \beq
 E_8  \supset SU(5)_{GUT}\times SU(5)_\perp \supset SU(5)_{GUT}\times U(1)_\perp^4~\cdot 
 \label{DP0}
 \eeq
 With respect to $t_i$ notation, the non-trivial $SU(5)_{GUT}$ representations given in~(\ref{DP00}) are designated as 
  $\bar 5_{t_i+t_j}, 10_{t_i}$ and the $SU(5)_{GUT}$ singlets  emerging from $SU(5)_{\perp}$ 
  adjoint decomposition are denoted with $ 1_{t_i-t_j}\equiv \theta_{ij}$.
  This way, the $U(1)_{\perp}^4$ invariance of each superpotential term is ensured  as soon as
 the tracelessness condition  $\sum_{i=1}^5t_i=0$ is satisfied~\footnote{\it 
 In the geometric language, we say that  the 10-plets and 5-plets  are found in the intersections
 of the  $SU(5)_{GUT}$ divisor with 7-branes associated with the $U(1)_{\perp}$'s, usually called matter curves and 
 designated as $\Sigma_{10_{t_i}}$ and $ \Sigma_{5_{t_i+t_j}}$.}.

 The second important available toolkit  in F-theory model building includes  the various fluxes  which  are turned on along 
 the various abelian factors and can be used for the following tasks.  
 Firstly,  appropriate fluxes are  introduced for the breaking symmetry mechanism and can be regarded as the surrogate 
 of the vacuum expectation value (vev) of the Higgs field, particularly in the case of manifold geometries 
 (such as del Pezzo surfaces)  where the latter is absent from the massless spectrum. 
  The $SU(5)_{GUT}$ symmetry in particular, breaks to the Standard Model  with a hypercharge flux. Secondly, fluxes are 
  used to generate the observed chirality  of the massless spectrum. In order to describe their implications in the present 
  construction we distinguish them into two classes. Initially, a flux is introduced along a $U(1)_{\perp}$ and
 its geometric restriction along a specific matter  curve $\Sigma_{n_j}$ is parametrised with
 an integer $M_{n_j}$. Then, the  chiralities  of the $SU(5)$ representations are given by 
 \bea
  \# 5_i-\# \overline{5}_i&=& M_{5_i}\label{M5ii}\\
 \# 10_j-\# \overline{10}_j&=& M_{10_j}\label{M10jj}
 \eea 
The hypercharge flux introduced to break $SU(5)_{GUT}$ is also responsible for the splitting of the 
$SU(5)$ representations.  If some integers   $N_{i,j}$ represent hyperfluxes piercing  
certain  matter curves, the 10-plets and 5-plets  split according to:
\bea
{10}_{t_{j}}=
\left\{\begin{array}{ll}
	n_{{(3,2)}_{\frac 16}}-n_{{(\bar 3,2)}_{-\frac 16}}&=\;M_{10_j}\\
	n_{{(\bar 3,1)}_{-\frac 23}}-n_{{(
			3,1)}_{\frac 23}}&=\;M_{10_j}-N_j\\
	n_{(1,1)_{+1}}-n_{(1,1)_{-1}}& =\;M_{10_j}+N_j\\
\end{array}\right.\,,
\label{F10j}
\eea
\bea
{5}_{t_{i}}=
\left\{\begin{array}{ll}
	n_{(3,1)_{-\frac 13}}-n_{(\bar{3},1)_{+\frac 13}}&=\;M_{5_i}\\
	n_{(1,2)_{+\frac 12}}-n_{(1,2)_{-\frac 12}}& =\;M_{5_i}+N_i\,\cdot\\
\end{array}\right.
\label{F5i}
\eea

The splitting of $SU(5)_{GUT}$ representations has  important and far reaching implications 
in model building.  With a suitable choice of  the flux integers on  the Higgs curve(s), $\Sigma_{5_{h_u}},\Sigma_{5_{h_d}}$,
one can generate the doublet-triplet splitting of the Higgs 5-plets and suppress dimension-five
operators contributing to proton decay.  We note in passing that this is a novel feature of F-theory  and there is no
counterpart in ordinary GUTs.
 Another important observation is that, by  virtue of the hypercharge flux, 
the  SM representations of the same family may no logner be components of the same 5-plet.
As a result, the $U(1)_{\perp}$ charges may be different not only between various families,
but even within SM fields of the same fermion generation.  It is exacly this property that is 
expected to allow the existence of non-universal flavour couplings of the corresponding $Z'$
boson to lepton fields.  Therefore, we conclude that the hyperflux splitting mechansim provides a
 natural way to implement the idea of a new neutral $Z'$  gauge boson 
coupled differently either to the third- or to a vector-like  family within the F-theory motivated framework.

\section{An $SU(5)$ model}
Having discussed the essential features of F-GUT model building, we now proceed in a specific model.
We have seen already that, in principle, there are at most four independent roots $t_i$.  
On the other hand, the four Cartan generators corresponding to  $U(1)_\perp^4$ are expressed as:
\bea 
H_1&=&\frac{1}{2}{\rm diag}(1,-1,0,0,0),\nn \\
H_2&=&\frac{1}{2\sqrt{3}}{\rm diag}(1,1,-2,0,0),\nn\\
H_3&=&\frac{1}{2\sqrt{6}}{\rm diag}(1,1,1,-3,0),\label{Hi}\\
H_4&=&\frac{1}{2\sqrt{10}}{\rm diag}(1,1,1,1,-4).\nn
\eea 
Throughout this paper we shall assume the minimal  ${\cal Z}_2$
 monodromy~\footnote{In this semi-local approach, several essential features of the effective field
theory model are determined from the topological properties  of the associated spectral surface, ${\cal C}$,
described by a fifth degree polynomial $
{\cal C}:\; \sum_{k=0}^5 b_k t^{5-k}=0~.$
The essential ingredients  specifying these properties are the coefficients $b_k$ 
with well defined topological properties.  The roots of the
polynomial determine the  weight vectors $t_{1,...,5}$.
In general, however, depending on the properties of the specific compact manifold, the solutions 
$t_i=t_i(b_k)$'s  imply that there is an action on the roots  $t_i$ of a non-trivial monodromy group which is
a subgroup of the Weyl group $W(SU(5)_{\perp})=S_5$. Such subgroups are the alternating
groups ${\cal A}_n$, the dihedral groups ${\cal D}_n$ and cyclic groups ${\cal Z}_n$, $n\le 5$ and the
Klein four-group ${\cal Z}_2\times {\cal Z}_2$.
}
\bea
{\cal  Z}_2\; {\rm monodromy}:  t_1\leftrightarrow t_2\label{Z2M}
\eea 
As a consequnce, this action reduces the number of the abelian factors by one, $U(1)_{\perp}^4\to U(1)_{\perp}^3$,
and allows a tree-level Yukawa coupling for the top-quark. 

For the rest of our analysis,  it is of interest to consider the following sequence of flux breaking, which may be 
associated with different scales
\bea
E_8
& \supset & E_6\times U(1)_\perp^2 \label{E60} \\
& \supset & SO(10)\times U(1)_{\psi}\times U(1)_\perp^2 \label{SO(10)} \\
& \supset & SU(5)_{GUT} \times  U(1)_{\chi}  \times U(1)_{\psi} \times  U(1)_\perp^2 . \label{SU(5)0}
\eea
Furthermore, as described above, we assume a $U(1)_Y$ flux which realises the $ SU(5)_{GUT}$ breaking and at the same time triggers the 
 doublet-triplet splitting and generates the chiral familes of the Standard Model through~(\ref{F10j}) and (\ref{F5i}).

It is convenient to choose a basis for the weight vectors such that the charge generators have the form
 where $Q_{\perp}$ is the charge of the $U(1)_{\perp}$ in the breaking pattern of Eq. (\ref{E60}) that 
 remains after imposing the $t_1\leftrightarrow t_2$ monodromy.
 In the conventional basis for the  $SU(5)_\perp$ generators in Eq. (\ref{Hi})
 the unbroken generators are identified as follows:
\beq
 H_2=Q_{\perp}, \ \ H_3 = Q_{\psi}, \ \ H_4 = -Q_{\chi}.
\eeq
This almost trivial equivalence shows that the $SU(5)_{GUT}$ states in Eq. (\ref{DP00})
have well defined $E_6$ charges $Q_{\chi}$ and $Q_{\psi}$.
For example $SU(5)$ singlets will in general carry $Q_{\chi}$ and $Q_{\psi}$ charges
which originate from $E_6$ and which may be unbroken. The equivalence will provide
insights into both anomaly cancellation and the origin of $R$-parity for example,
in terms of the underlying $E_6$ structure, in the explicit models discussed later.

In order to work consistently, we use the normalised $SU(5)_\perp$ generators

\bea
Q_\perp &= \frac{1}{2\sqrt{3}}\diag{1,1,-2,0,0}\\
Q_\psi &= \frac{1}{2\sqrt{6}}\diag{1,1,1,-3,0}\\
Q_\chi & = \frac{1}{2\sqrt{10}}\diag{1,1,1,1,-4}
\eea
and the low energy $U(1)^\prime$ is generated by
\beq
Q^\prime = c_1 Q_\perp + c_2 Q_\psi + c_3 Q_\chi\label{Qprime}
\eeq
which, in order to retain $SU(5)_\perp$ normalisation, are subject to the condition
\beq
c_1^2+c_2^2+c_3^2 =1
\eeq

In order to proceed, we need to determine the spectrum and assign the corresponding $Q^\prime$ charges for all fields.
We first remark that, in order to obtain three chiral families, the set of  integers $M_{n_j}$ associated with the $U(1)_{\perp}$ 
flux in formulae (\ref{M10jj}, \ref{M5ii})  must obay 
\bea
\sum_i M_{5_i} =-\sum_j M_{10_j} &= -3 \label{GeoC}
\eea
This automatically implies also that the total flux vanishes, $\sum M_{5_i} + \sum M_{10_j} = 0 $.
Furthermore, the $SU(5)$ representations  are localised on matter curves  possessing certain topological characteristics 
which  determine the various properties of the effective SM theory. Important r\^ole, in particular,  is played
 by the homologies of these curves which  are determined in accordance with the ${\cal Z}_2$ factorisation of the spectral cover 
 polynomial.  It turns out that the homologies of the whole set of matter curves can be 
expressed in terms of only three arbitrary parameters $\chi_{6,7,8}$~\cite{Dudas:2010zb}.
 If  ${\cal F}_Y$ represents the hypercharge flux, its `dot' product with $\chi_{6,7,8}$ determines
  three   hyperflux  integers
\bea
N_7={\cal F}_Y\cdot \chi_7,\;   N_8={\cal F}_Y\cdot \chi_8,\;   N_9={\cal F}_Y\cdot \chi_9\ ,
\eea
and we define $\tilde N = N_7 + N_8 +N_9$.
In conclusion,  multiplicities of the SM states are determined by  $N_i, i=7,8,9$
 associated with hyperflux restrictions and the $U(1)_{\perp}$ flux integers $M_{n_j}$.
 More specifically, for a given choice of these numbers  we can determine the exact number of the SM states
 residing on each matter curve  using the formulae (\ref{F10j},\ref{F5i})
 and then apply the definition (\ref{Qprime}) to find their corresponding charges. We refer to
 previous work for the details~\cite{Callaghan:2011jj}, while here,  we only  present 
selected properties of the spectrum in Table~\ref{T1}.

\begin{table}[H]
	\small
	\centerline{
		\begin{tabular}{|c|c|c|c|c|}
			\hline
			Curve Name   &           $Q^\prime$        &    $N_Y$    &       M       &        SM Content      
			\\ \hline
			$\Sigma_{5_{H_u (-2t_1)}}$   &     $-\frac{c_1}{\sqrt{3}}-\frac{c_2}{\sqrt{6}}-\frac{c_3}{\sqrt{10}}$     & $\tilde N$  & $M_{5_{H_u}}$ & $M_{5_{H_u}}\overline{d^c} + (M_{5_{H_u}}+\tilde N)\overline{L}$ 
			\\
			$\Sigma_{5_{1,\pm (t_1+t_3)}} $ &    $\frac{5 \sqrt{3} c_1-5 \sqrt{6} c_2-3 \sqrt{10} c_3}{30} $    & $-\tilde N$ &  $M_{5_{1}}$  &     $M_{5_1}\overline{d^c} + (M_{5_1}-\tilde N)\overline{L}$    
			\\
			$\Sigma_{5_{2,\pm (t_1+t_4)}}$  &    $-\frac{c_1}{2 \sqrt{3}}+\frac{c_2}{\sqrt{6}}-\frac{c_3}{\sqrt{10}}$     & $-\tilde N$ &  $M_{5_{2}}$  &     $M_{5_2}\overline{d^c} + (M_{5_2}-\tilde N)\overline{L}$    
			\\
			$\Sigma_{5_{3,\pm (t_1+t_5)}}$ & $\frac{-10 \sqrt{3} c_1-5 \sqrt{6} c_2+9 \sqrt{10} c_3}{60} $ & $-\tilde N$ &  $M_{5_{3}}$  &     $M_{5_3}\overline{d^c} + (M_{5_3}-\tilde N)\overline{L}$    
			\\
			$\Sigma_{5_{4,\pm (t_3+t_4)}} $ &     $\frac{c_1}{\sqrt{3}}+\frac{c_2}{\sqrt{6}}-\frac{c_3}{\sqrt{10}}$      &  $N_7+N_8$  &  $M_{5_{4}}$  &     $M_{5_4}\overline{d^c} + (M_{5_4}+N_7+N_8)\overline{L}$      
			\\
			$\Sigma_{5_{5,\pm (t_3+t_5)}}$ &  $\frac{20 \sqrt{3} c_1-5 \sqrt{6} c_2+9 \sqrt{10} c_3}{60} $  &  $N_7+N_9$  &  $M_{5_{5}}$  &     $M_{5_5}\overline{d^c} + (M_{5_5}+N_7+N_9)\overline{L}$      
			\\
			$\Sigma_{5_{6,\pm (t_4+t_5)}}$ &           $\frac{5 \sqrt{6} c_2+3 \sqrt{10} c_3}{20}$            &  $N_8+N_9$  &  $M_{5_{6}}$  &     $M_{5_6}\overline{d^c} + (M_{5_6}+N_8+N_9)\overline{L}$      
			\\ \hline
			$\Sigma_{10_{t,\pm t_1}}$    &  $\frac{10 \sqrt{3} c_1+5 \sqrt{6} c_2+3 \sqrt{10} c_3}{60}$   & $-\tilde N$ & $M_{10_{t}}$  &   $M_{10_t}Q+(M_{10_t}+\tilde N) u^c +(M_{10_t}-\tilde N)e^c$   
			\\
			$\Sigma_{10_{2,\pm t_3}}$   &   $\frac{-20 \sqrt{3} c_1+5 \sqrt{6} c_2+3 \sqrt{10} c_3}{60}$    &    $N_7$    & $M_{10_{2}}$  &       $M_{10_2}Q+(M_{10_2}- N_7) u^c +(M_{10_2}+ N_7)e^c$        
			\\
			$\Sigma_{10_{3,\pm t_4}}$    &            $\frac{\sqrt{10} c_3-5 \sqrt{6} c_2}{20} $  &    $N_8$    & $M_{10_{3}}$  &   $M_{10_3}Q+(M_{10_3}- N_8) u^c +(M_{10_3}+N_8)e^c$       
			\\
			$\Sigma_{10_{4,\pm t_5}}$     &     $-\sqrt{\frac{2}{5}} c_3$   &    $N_9$    & $M_{10_{4}}$  &   $M_{10_4}Q+(M_{10_4}- N_9) u^c +(M_{10_4}+N_9)e^c$       
			\\ \hline
	\end{tabular}}
\caption{ The first column shows the matter curve accommodating the  $SU(5)$ representations, and the $U(1)_{\perp}$ weights  ($\pm$ refer to $10/\overline{10}$ and $\bar 5/5$ respectively).
	The second column  displays their $Q'$  charges and columns 3 and 4 the $U(1)_{Y,\perp}$ 
	flux integers $N_i, M_{n_j}$. The last column shows the SM multiplicities in terms of the $N_i, M_{n_j}$ numbers. Singlet 
fields $\theta_{ij}$ are not presented in this Table (see however text of section 2.)}
\label{T1}
\end{table}


\subsection{Anomalies} 

The presence of an additional $U(1)'$ symmetry in the effective model is associated with  contributions to cubic and mixed anomalies with the SM gauge group $G_{SM}$. On the other hand,  we have seen that the multiplicities in the SM spectrum are determined by the flux parameters $M_{n_i}, N_j$ and therefore these numbers  are directly  involved  in the anomaly constraints. These involve the $U(1)^\prime$ anomalies with the $G_{SM}$ currents  as well as 
quartic and trace anomaly conditions.
 Thus, the flux integers $M_{n_i}, N_j$ are subject to certain restrictions. 
 All anomalies were systematically computed using the SUSYNO package \cite{Fonseca:2011sy}, those involving $G_{SM}$ currents in particular  are as follows:
\beq
\small
\begin{split}
	\m{A}_{G_{SM}^2 \times U(1)^\prime} = &{\tiny \left(-\frac{c_1}{2 \sqrt{3}}-\frac{c_2}{2 \sqrt{6}}-\frac{c_3}{2 \sqrt{10}}\right) M_{5_{H_u}}+\left(\frac{\sqrt{3} c_1}{4}+\frac{1}{4} \sqrt{\frac{3}{2}} c_2+\frac{3 c_3}{4 \sqrt{10}}\right) M_{10_t}+\left(\frac{c_1}{4\sqrt{3}}-\frac{c_2}{2 \sqrt{6}}-\frac{c_3}{2 \sqrt{10}}\right) M_{5_1}}\\
	  &+{\tiny \left(-\frac{c_1}{4 \sqrt{3}}+\frac{c_2}{2 \sqrt{6}}-\frac{c_3}{2 \sqrt{10}}\right) M_{5_2}+\left(-\frac{c_1}{4 \sqrt{3}}-\frac{c_2}{4\sqrt{6}}+\frac{3 c_3}{4 \sqrt{10}}\right) M_{5_3}+\left(\frac{c_1}{2 \sqrt{3}}+\frac{c_2}{2 \sqrt{6}}-\frac{c_3}{2 \sqrt{10}}\right) M_{5_4}}\\
	  &+{\tiny \left(\frac{c_1}{2 \sqrt{3}}-\frac{c_2}{4 \sqrt{6}}+\frac{3 c_3}{4\sqrt{10}}\right) M_{5_5}+\left(\frac{1}{4} \sqrt{\frac{3}{2}} c_2+\frac{3 c_3}{4 \sqrt{10}}\right) M_{5_6}+\left(-\frac{1}{2} \sqrt{3} c_1+\frac{1}{4} \sqrt{\frac{3}{2}} c_2+\frac{3 c_3}{4 \sqrt{10}}\right)M_{10_2}}\\
	  &+{\tiny \left(\frac{3 c_3}{4 \sqrt{10}}-\frac{3}{4} \sqrt{\frac{3}{2}} c_2\right) M_{10_3}-\frac{3 c_3 }{\sqrt{10}}M_{10_4}+\frac{1}{4} \sqrt{3} c_1 N_7+\left(\frac{c_1}{4 \sqrt{3}}+\frac{c_2}{\sqrt{6}}\right) N_8}+{\tiny \left(\frac{c_1}{4 \sqrt{3}}+\frac{c_2}{4 \sqrt{6}}+\frac{1}{4} \sqrt{\frac{5}{2}} c_3\right) N_9} \\
	  \m{A}_{U(1)_Y \times U(1)^{\prime 2}}   =&{\tiny\frac{3}{2} c_1^2 N_7+\left(\frac{c_1^2}{6}+\frac{2}{3} \sqrt{2} c_2 c_1+\frac{4 c_2^2}{3}\right) N_8+\left(\frac{c_1^2}{6}+\frac{c_2 c_1}{3 \sqrt{2}}+\sqrt{\frac{5}{6}} c_3 c_1+\frac{c_2^2}{12}+\frac{5c_3^2}{4}+\frac{1}{2} \sqrt{\frac{5}{3}} c_2 c_3\right) N_9 }\\
	  \m{A}_{G_{SM}^3}  =& {\tiny \frac{5}{36} \left( M_{5_{H_u}}+ M_{10_t}+ M_{5_1}+ M_{5_2}+M_{5_3}+M_{5_4}+ M_{5_5}+M_{5_6}+ M_{10_2}+M_{10_3}+ M_{10_4} \right)} \ .
\end{split}
\eeq

In addition, we have to consider the trace anomaly, which is related to the D-flatness constraint and the associated Green-Schwarz cancellation mechanism, and the cubic anomaly that involve  
contributions from all singlet fields $\theta_{ij}$.  To avoid clutter in this short note, these are not included in the above equations, however,  they  can be computed in a straightforword manner and are taken into account in our subsequent computations. 
 We should point out however, that in order to have a sizable effect in anomalous $B$-decays,  the solution of the anomaly cancellation equations must be compatible with a TeV scale  mass of the corresponding $Z'$ boson.

Hence, our strategy is as follows:  we start by working out  the solutions with the anomalies involving $SM$ currents, 
$G_{SM}^2\times U(1)^\prime$ and $U(1)_Y\times U(1)^{\prime2}$ 
and solve for all $c_i$ coefficients  subject to the normalization condition $c_1^2+c_2^2+c_3^2=1$. This will provide a solution $c_i=c_i(M_{n_j},N_k)$.
First, we notice that the MSSM chirality trivially solves for $\m{A}_{G_{SM}^3}=0$. 
 Then, we check whether there is a solution for $U(1)^{\prime 3}$, $\mbox{Tr }Q^\prime$ anomalies using the $c_i$ derived before, keeping only the solutions for which the charges can be written in a single radical form.\footnote{This means that the charges for each state can all factor out $\sqrt{a}$, where $a$ is a rational number, such that it can be absorbed into the definition of the $Z^\prime$ coupling constant and hence making all charges appear rational numbers themselves.} 
 These last two constraints will be solved for multiplicities of singlet fields $\theta_{ij}$. And in our search for anomaly free solutions, we will seek models with the MSSM spectrum, as well as icluding those with complete vector-like families. 

\section{Minimal MSSM type spectrum but with a non-universal $Z'$}

We start with models that have the same spectrum of the MSSM with the gauge group extended by a single $U(1)^\prime$ factor.
In order to have the MSSM chirality, the flux data must respect the geometric constraints~(\ref{GeoC})
which are used to solve for $M_{5_6}$ and $M_{10_4}$. Notice that this solves $G_{SM}^3$ anomaly trivially. Further, the MSSM spectrum was imposed
\bea
\# L + \# \overline L = 5 \\
\# d^c = \# Q = \# u^c = \# e^c = 3
\eea
and we demanded the existence of a solution for the doublet-triplet splitting problem
\beq
|N_7| + |N_8| + |N_9| \neq 0 \ .
\eeq
Since every state inside the same matter curve shares the same charge $Q^\prime$, a non-universal model will require the families to be supported in different matter curves.
In order to still scan models that allow for that to happen either only for the $\Sigma_{5}$ or $\Sigma_{10}$ matter curves, we allow the flux parameters to vary between
 \beq
 N_i, \ M_{n_j} \in [-3,3] \ .
 \eeq
In order to guarantee a renormalisable top quark Yukawa coupling, we identify $H_u\in 5_{-2t_1}$   and  $t\in 10_{t_1}$
as these states are involved in the renormalisable Yukawa interaction
\beq
10_{t_1}10_{t_1}5_{-2t_1}\equiv 10_t 10_t5_{H_u} ~\cdot\label{Ytop}
\eeq
This accounts for
\bea
	M_{5_{H_u}}+ \tilde N \geq 1, \;
	M_{10_t } \geq 1, \;
	M_{10_t}+\tilde{N} \geq 1\,,
\eea
and we further assume $M_{10_t} = 1$ to fix the left-handed top $t_L$.

We also restrict our searches to solutions that have a single and isolated $H_u$ in the $5_{H_u}$ curve, as we are preventing exotics in the spectrum. This accounts for
\bea
	M_{5_{H_u}} = 0\,, \;
	\tilde N = 1\,,
\eea
where the last equation can be used to solve for one of the $N_i$ flux
\beq
N_7 = 1 - N_8 - N_9 \ .
\eeq

After scanning the entire parameter space, we find that there are only 48 solutions  respecting all the conditions above, which fall in just four different classes of models. We notice that these are the only classes of models for an $SU(5)$ GUT with a  $\m{Z}_2$ monodromy with a consistent $Z^\prime$ interaction. The models inside the same class differ as to how the states are distributed amongst different curves and the perpendicular charges they carry. The four distinct classes are, in the $G_{SM}\times U(1)^\prime$ basis
\bea
	\mbox{Class 1: }& (H_u)_{2\frac{1}{\sqrt{15}}}+(H_d)_{-2\frac{1}{\sqrt{15}}} + 3 \times \overline 5_{\frac{1}{2}\frac{1}{\sqrt{15}}}+2\times 10_{-\frac{1}{\sqrt{15}}}+10_{\frac{3}{2}\frac{1}{\sqrt{15}}} \label{eq:model1}\\
		\mbox{Class 2: }& (H_u)_{\frac{1}{2}\frac{1}{\sqrt{15}}}+(H_d)_{-\frac{1}{2}\frac{1}{\sqrt{15}}} + 2 \times \overline 5_{2\frac{1}{\sqrt{15}}}+\overline{5}_{-\frac{7}{4}\frac{1}{\sqrt{15}}}+3\times 10_{-\frac{1}{4}\frac{1}{\sqrt{15}}} \\
		\mbox{Class 3: } & (H_u)_{\frac{3}{2}\frac{1}{\sqrt{10}}}+\overline{5}_{-\frac{1}{4}\frac{1}{\sqrt{10}}}+\overline{5}_{\frac{1}{\sqrt{10}}}+(2 L+d^c)_{-\frac{3}{2}\frac{1}{\sqrt{10}}}+2 \times 10_{-\frac{3}{4}}+10_{\frac{7}{4}\frac{1}{\sqrt{10}}} \\
		\mbox{Class 4: }&(H_u)_{\frac{3}{2}\frac{1}{\sqrt{10}}}+(H_d)_{-\frac{3}{2}\frac{1}{\sqrt{10}}}+ \overline{5}_{-\frac{1}{4}\frac{1}{\sqrt{10}}}+\overline{5}_{\frac{1}{\sqrt{10}}}+\overline{5}_{\frac{9}{4}\frac{1}{\sqrt{10}}}+2 \times 10_{-\frac{3}{4}\frac{1}{\sqrt{10}}}+10_{\frac{1}{2}\frac{1}{\sqrt{10}}}
\eea
	in addition to $Q^\prime \to -Q^\prime$. 

	In Table \ref{tab:datamodels1to4} we present an explicit element of each class, and the distribution of the states amongst different matter curves is presented in Table \ref{tab:models1to4}. 

	\begin{table}[H]
		\small
		\centerline{
			\begin{tabular}{|c|c|c|c|c|c|c|c|c|c|c|c|c|c|c|c|c|c|}
				\hline
				Model & $c_1$                 & $c_2$                            & $c_3$                            & $M_{5_{H_u}}$ & $M_{5_1}$ & $M_{5_2}$ & $M_{5_3}$ & $M_{5_4}$ & $M_{5_5}$ & $M_{5_6}$ & $M_{10_t}$ & $M_{10_2}$ & $M_{10_3}$ & $M_{10_4}$ & $N_7$ & $N_8$ & $N_9$ \\ \hline
				$1$   & $-\frac{\sqrt{5}}{3}$ & $\frac{1}{6}\sqrt{\frac{5}{2}}$  & $-\frac{1}{2}\sqrt{\frac{3}{2}}$ & $0$           & $0$       & $0$       & $-1$      & $-1$      & $0$       & $-1$      & $1$        & $1$        & $1$        & $0$        & $0$   & $1$   & $0$   \\ 
				$2$   & $-\frac{\sqrt{5}}{3}$ & $-\frac{1}{3}\sqrt{\frac{5}{2}}$ & $\frac{1}{\sqrt{6}}$             & $0$           & $0$       & $-1$      & $0$       & $0$       & $-1$      & $-1$      & $1$        & $1$        & $0$        & $1$        & $0$   & $0$   & $1$   \\ 
				$3$   & $-\sqrt{\frac{5}{6}}$ & $-\frac{1}{8}\sqrt{\frac{5}{3}}$ & $\frac{3}{8}$                    & $0$           & $0$       & $0$       & $-1$      & $0$       & $-1$      & $-1$      & $1$        & $1$        & $0$        & $1$        & $0$   & $0$   & $1$   \\ 
				$4$   & $-\sqrt{\frac{5}{6}}$ & $\frac{1}{4}\sqrt{\frac{5}{3}}$  & $-\frac{1}{4}$                   & $0$           & $0$       & $0$       & $0$       & $-1$      & $-1$      & $-1$      & $1$        & $0$        & $1$        & $1$        & $0$   & $1$   & $0$   \\ \hline
				\end{tabular}
		}
		\caption{Flux data and $c_i$ coefficients for explicit models 1 to 4 of the classes 1 to 4, respectively}
		\label{tab:datamodels1to4}
	\end{table}

	As the left-handed top-quark is always identified as being in $10_t$, we notice that in models $1,3$ and $4$ the first and second family left-handed quarks have different $U(1)^\prime$ charges, which can be checked for all elements of the respective classes. As a result, the $Z^\prime$ mass for these models is constrained to be $m_{Z^\prime}\gtrsim 10^5$ TeV
	or otherwise $Z^\prime$ would contribute too strongly at tree-level to $K^0-\overline K^0$ oscillations. In addition, model $2$ leaves a certain ambiguity about what $L$ state should be identified as $H_d$.

	\begin{table}[H]
		\small
		\centerline{
				\begin{tabular}{cc|c|c||c|c||c|c||c|c|}
				\cline{3-10}
				\multicolumn{1}{c}{}             & \multicolumn{1}{c|}{} & \multicolumn{2}{c||}{Model 1}           & \multicolumn{2}{c||}{Model 2}           & \multicolumn{2}{c||}{Model 3}       & \multicolumn{2}{c|}{Model 4}       \\ \hline
				\multicolumn{1}{|c|}{Curve} & Weights               & $Q^\prime \sqrt{15}$ & SM Content      & $Q^\prime \sqrt{15}$ & SM Content      & $Q^\prime \sqrt{10}$ & SM Content  & $Q^\prime \sqrt{10}$ & SM Content  \\ \hline
				\multicolumn{1}{|c|}{$5_{H_u}$}  & $-2t_1$               & $2$                  & $H_u$           & $\frac{1}{2}$        & $H_u$           & $\frac{3}{2}$        & $H_u$       & $\frac{3}{2}$        & $H_u$       \\
				\multicolumn{1}{|c|}{$5_1$}      & $t_1+t_3$             & $\frac{1}{2}$        & $L$             & $ 2$                 & $d^c+2 L$       & $1$                  & $L$         & $1$                  & $L$         \\
				\multicolumn{1}{|c|}{$5_2$}      & $t_1+t_4 $            & $-2$                 & $H_d$           & $-\frac{1}{2}$       & $H_d$           & $-\frac{1}{4}$       & $L$         & $-\frac{3}{2}$       & $H_d$       \\
				\multicolumn{1}{|c|}{$5_3$}      & $t_1+t_5 $            & $\frac{1}{2 }$       & ${d^c} + 2 {L}$ & $-\frac{7}{4}$       & $L$             & $-\frac{3}{2}$       & $d^c+2L$    & $-\frac{1}{4}$       & $L$         \\
				\multicolumn{1}{|c|}{$5_4$}      & $t_3+t_4 $            & $\frac{1}{2 }$       & ${d^c}$         & $2$                  & $d^c$           & $-\frac{9}{4}$       & $-$         & $1$                  & $d^c$       \\
				\multicolumn{1}{|c|}{$5_5$}      & $t_3+t_5 $            & $-3$                 & $-$             & $-\frac{3}{4}$       & $-$             & $1$                  & $d^c$       & $ \frac{9}{4}$       & $d^c+L$     \\
				\multicolumn{1}{|c|}{$5_6$}      & $t_4+t_5 $            & $\frac{1}{2 }$       & ${d^c}$         & $-\frac{7}{4}$       & $d^c$           & $-\frac{1}{4}$       & $d^c$       & $-\frac{1}{4}$       & $d^c$       \\ \hline
				\multicolumn{1}{|c|}{$10_t$}     & $t_1$                 & $-1$                 & $Q+2 u^c+e^c$   & $-\frac{1}{4}$       & $Q+2 u^c$       & $-\frac{3}{4}$       & $Q+2 u^c$   & $-\frac{3}{4}$       & $Q+2 u^c$   \\
				\multicolumn{1}{|c|}{$10_2$}     & $t_3$                 & $\frac{3}{2}$        & $Q+u^c +e^c$    & $\frac{9}{4}$        & $-$             & $\frac{7}{4}$        & $Q+u^c+e^c$ & $\frac{7}{4}$        & $-$         \\
				\multicolumn{1}{|c|}{$10_3$}     & $t_4$                 & $-1$                 & $Q+2e^c$        & $-\frac{1}{4}$       & $2 Q+u^c+3 e^c$ & $\frac{1}{2}$        & $-$         & $-\frac{3}{4}$       & $Q+2 e^c$   \\
				\multicolumn{1}{|c|}{$10_4$}     & $t_5$                 & $\frac{3}{2}$        & $-$             & $-\frac{3}{2}$       & $-$             & $-\frac{3}{4}$       & $Q+2 e^c$   & $ \frac{1}{2}$       & $Q+u^c+e^c$ \\ \hline
				\end{tabular}	
		}
		\caption{Explicit Models 1 to 4 using the data in Table \ref{tab:datamodels1to4}}
		\label{tab:models1to4}
		\end{table}

\section{Models with a vector-like family with non-universal couplings to a $Z'$}

Another interesting possibility is that there are new vector-like states in the spectrum such that the left-handed states have a different charge under $U(1)^\prime$. As shown in~\cite{King:2017anf}, this can lead to non-universality in the regular matter through mixing in the new neutral current coupling to $Z^\prime$.

In order to preserve unification, we need these vector-like states to form complete $SU(5)$ representations. In particular, we look 
for models that allow for a complete vector-like family. In addition, as we wish to preserve the top-quark assignment, the following 
simplifying assumptions were taken
\beq
	M_{5_{H_u}}=0~, \; 	\tilde N = 1~,
\eeq
in order to obtain models where the $H_u$ is isolated in its own matter curve, as can be seen from (\ref{F5i}). With this simplification, computational time is significantly reduced and we are able to probe the entire region of parameter space of interest~\footnote{Note that 
 in order for the extra  vector-like $({\bf 10}+\overline{\bf 5})+(\overline{\bf 10}+{\bf 5})$ to induce non-universality from mixing, 
it has to have different charges than regular matter, therefore we limit the values of flux data to be within $[-3,3]$.}.

There are 4067 models with one extra vector-like family, but only 397 classes when considering only $G_{SM}\times U(1)^\prime$ charge assignments. Here we present three of these classes of models
\bea
	\mbox{Class 5: }&(H_u)_{4\frac{1}{\sqrt{85}}}+(H_d)_{-4\frac{1}{\sqrt{85}}}+3\times \overline{5}_{\frac{7}{2}\frac{1}{\sqrt{85}}}+\overline{5}_{\frac{3}{2}\frac{1}{\sqrt{85}}}+5_{6\frac{1}{\sqrt{85}}}+3 \times 10_{2\frac{1}{\sqrt{85}}}+10_{-\frac{11}{2}\frac{1}{\sqrt{85}}}+\overline{10}_{\frac{1}{2}\frac{1}{\sqrt{85}}} \\
	\mbox{Class 6: }&(H_u)_{\frac{3}{2}\frac{1}{\sqrt{10}}}+(H_d)_{-\frac{3}{2}\frac{1}{\sqrt{10}}}+\overline{5}_{\frac{1}{\sqrt{10}}}+\overline{5}_{-\frac{1}{4}\frac{1}{\sqrt{10}}}+\overline{5}_{-\frac{9}{4}\frac{1}{\sqrt{10}}}+L_{\frac{1}{\sqrt{10}}}+d^c_{-\frac{1}{4}\frac{1}{\sqrt{10}}}+5_{-\frac{1}{\sqrt{10}}}+3\times 10_{-\frac{3}{4}\frac{1}{\sqrt{10}}}\nonumber \\
	&+10_{\frac{7}{4}\frac{1}{\sqrt{10}}}+\overline{10}_{-\frac{1}{2}\frac{1}{\sqrt{10}}}\\
	\mbox{Class 7: }&(H_u)_{-\frac{1}{2}}+(H_d)_{\frac{1}{2}}+3\times \overline{5}_{-\frac{1}{4}}+\overline{5}_{\frac{3}{4}}+5_0+3\times 10_{\frac{1}{4}}+10_{-\frac{1}{2}}+\overline{10}_{\frac{1}{4}}~,
\eea
and we notice the the first and last ones provide models where regular matter is universal under the new interaction, while the non-universality is strickly induced by the extra vector-like states. As $K^0-\bar K^0$ mixing imposes strong constraints on
 non-universality in the first two families, these two cases, in principle, could be consistent with a light (few TeV) $Z'$ boson mass 
and  could be sufficient to explain the $B$-decay anomaly. On the contrary, models from the Class $6$ will probably require a heavy $Z^\prime$, as the first and second family right-handed down-quarks have different charges. 

In Table \ref{tab:dataex1to3} we present the flux data and solutions for the coefficients $c_i$ for three explicit
 realisations  of models respecting these conditions. The spectrum distribution amongst curves of these three models
 can be seen in  Table~\ref{tab:ex1to3}.

\begin{table}[H]
	\small
	\centerline{
		\begin{tabular}{|c|c|c|c|c|c|c|c|c|c|c|c|c|c|c|c|c|c|}
			\hline
			Model & $c_1$                 & $c_2$                            & $c_3$                            & $M_{5_{H_u}}$ & $M_{5_1}$ & $M_{5_2}$ & $M_{5_3}$ & $M_{5_4}$ & $M_{5_5}$ & $M_{5_6}$ & $M_{10_t}$ & $M_{10_2}$ & $M_{10_3}$ & $M_{10_4}$ & $N_7$ & $N_8$ & $N_9$ \\ \hline
			$5$ & $0$ & $\frac{1}{2}\sqrt{\frac{15}{34}}$ & $\frac{11}{2 \sqrt{34}}$ & $0$ & $0$ & $0$ & $0$ & $-1$ & $-3$ & $1$ & $1$ & $2$ & $ -1$ & $1$ & $1$ & $0$ & $0$ \\
			$6$ & $-\frac{\sqrt{\frac{5}{6}}}{2}$ & $-\frac{5 \sqrt{\frac{5}{3}}}{8}$ & $\frac{3}{8}$ & $0$ & $1$ & $-1$ & $0$ & $-1$ & $-2$ & $0$ & $2$ & $-1$ & $1$ & $1$ & $0$ & $0$ & $1$ \\
			$7$ & $\frac{\sqrt{3}}{2}$ & $-\frac{1}{4}\sqrt{\frac{3}{2}}$ & $\frac{1}{4}\sqrt{\frac{5}{2}}$ & $0$ & $0$ & $0$ & $1$ & $-3$ & $-1$ & $0$ & $2$ & $1$ & $1$ & $-1$ & $0$ & $1$ & $0$ \\\hline
			\end{tabular}
	}
	\caption{Flux data and $c_i$ coefficients for explicit models 5 to 7 of the classes 5 to 7, respectively}
	\label{tab:dataex1to3}
\end{table}

\begin{table}[H]
	\small
	\centering
	\begin{tabular}{cc|c|c||c|c||c|c|}
		\cline{3-8}
														 &             & \multicolumn{2}{c||}{Model 5}                                      & \multicolumn{2}{c||}{Model 6}                                      & \multicolumn{2}{c|}{Model 7}                                 \\ \hline
		\multicolumn{1}{|c|}{Curve}                      & Weights     & $Q^\prime\sqrt{85}$ & SM Content                                    & $Q^\prime\sqrt{10}$ & SM Content                                    & $Q^\prime$     & SM Content                                    \\ \hline
		\multicolumn{1}{|c|}{\multirow{2}{*}{$5_{H_u}$}} & $-2t_1$     & $-4$                & $H_u$                                         & $\frac{3}{2}$       & $H_u$                                         & $-\frac{1}{2}$ & $H_u$                                         \\
		\multicolumn{1}{|c|}{}                           & $2t_1$      & ---                 & ---                                           & ---                 & ---                                           & ---            & ---                                           \\
		\multicolumn{1}{|c|}{\multirow{2}{*}{$5_1$}}     & $-t_1-t_3$  & ---                 & ---                                           & $\frac{1}{4}$       & $\overline{d^c}$                              & ---            & ---                                           \\
		\multicolumn{1}{|c|}{}                           & $t_1+t_3$   & $4$                 & $H_d$                                         & ---                 & ---                                           & $-\frac{1}{4}$ & $L$                                           \\
		\multicolumn{1}{|c|}{\multirow{2}{*}{$5_2$}}     & $-t_1-t_4 $ & ---                 & ---                                           & ---                 & ---                                           & ---            & ---                                           \\
		\multicolumn{1}{|c|}{}                           & $t_1+t_4 $  & $\frac{3}{2}$       & $L$                                           & $1$                 & $d^c+2 L$                                     & $\frac{1}{2}$  & $H_d$                                         \\
		\multicolumn{1}{|c|}{\multirow{2}{*}{$5_3$}}     & $-t_1-t_5 $ & ---                 & ---                                           & ---                 & ---                                           & $0$            & $\overline{d^c}$                              \\
		\multicolumn{1}{|c|}{}                           & $t_1+t_5 $  & $-\frac{7}{2}$      & $L$                                           & $-\frac{3}{2}$      & $H_d$                                         & ---            & ---                                           \\
		\multicolumn{1}{|c|}{\multirow{2}{*}{$5_4$}}     & $-t_3-t_4 $ & ---                 & ---                                           & ---                 & ---                                           & ---            & ---                                           \\
		\multicolumn{1}{|c|}{}                           & $t_3+t_4 $  & $\frac{3}{2}$       & $d^c$                                         & $\frac{9}{4}$       & $d^c+L$                                       & $-\frac{1}{4}$ & $3 d^c + 2 L$                                 \\
		\multicolumn{1}{|c|}{\multirow{2}{*}{$5_5$}}     & $-t_3-t_5 $ & ---                 & ---                                           & ---                 & ---                                           & ---            & ---                                           \\
		\multicolumn{1}{|c|}{}                           & $t_3+t_5 $  & $-\frac{7}{2}$      & $3 d^c+2L$                                    & $-\frac{1}{4}$      & $2 d^c + L$                                   & $-\frac{3}{4}$ & $ {d^c}+ L$                                   \\
		\multicolumn{1}{|c|}{\multirow{2}{*}{$5_6$}}     & $-t_4-t_5 $ & $6$                 & $\overline{d^c}+\overline{L}$                 & $-1$                & $\overline{L}$                                & $0$            & $\overline{L}$                                \\
		\multicolumn{1}{|c|}{}                           & $t_4+t_5 $  & ---                 & ---                                           & ---                 & ---                                           & ---            & ---                                           \\ \hline
		\multicolumn{1}{|c|}{\multirow{2}{*}{$10_t$}}    & $t_1$       & $2$                 & $Q+2 u^c$                                     & $-\frac{3}{4}$      & $2 Q+3 u^c+e^c$                               & $\frac{1}{4}$  & $2Q+3 u^c+e^c $                               \\
		\multicolumn{1}{|c|}{}                           & $-t_1$      & ---                 & ---                                           & ---                 & ---                                           & ---            & ---                                           \\
		\multicolumn{1}{|c|}{\multirow{2}{*}{$10_2$}}    & $t_3$       & $2$                 & $2 Q+u^c+3 e^c$                               & ---                 & ---                                           & $-\frac{1}{2}$ & $Q + u^c+ e^c$                                \\
		\multicolumn{1}{|c|}{}                           & $-t_3$      & ---                 & ---                                           & $-\frac{1}{2}$      & $\overline{Q}+\overline{u^c}+ \overline{e^c}$ & ---            & ---                                           \\
		\multicolumn{1}{|c|}{\multirow{2}{*}{$10_3$}}    & $t_4$       & ---                 & ---                                           & $\frac{7}{4}$       & $Q+u^c+e^c$                                   & $\frac{1}{4}$  & $Q+2 e^c$                                     \\
		\multicolumn{1}{|c|}{}                           & $-t_4$      & $\frac{1}{2}$       & $\overline{Q}+\overline{u^c}+ \overline{e^c}$ & ---                 & ---                                           & ---            & ---                                           \\
		\multicolumn{1}{|c|}{\multirow{2}{*}{$10_4$}}    & $t_5$       & $\frac{11}{2}$      & $Q+u^c+e^c$                                   & $-\frac{3}{4}$      & $Q+2 e^c$                                     & ---            & ---                                           \\
		\multicolumn{1}{|c|}{}                           & $-t_5$      & ---                 & ---                                           & ---                 & ---                                           & $\frac{1}{4}$  & $\overline{Q}+\overline{u^c}+ \overline{e^c}$ \\ \hline
		\end{tabular}
	\caption{Explicit models 5 to 7 using the data in Table \ref{tab:dataex1to3}}
	\label{tab:ex1to3}
	\end{table}

\subsection{A brief discussion of model 7}

Next, we briefly discuss the main phenomenological properties of the emergent low energy effective theory  with 
the spectrum given by model 7 of Table~\ref{tab:ex1to3}. Because of the constraints imposed by the  $K^0-\bar K^0$ 
oscillations, as discussed above,  the  left-handed quark doublets of the  three chiral fermion families should be 
accommodated in $10_t, 10_3$ which have equal charges, $Q'=\frac 14$, with respect to  $U(1)'$. Similarly, the 
three down quarks and lepton doublets are distributed in the fiveplets $\bar 5_1, \bar 5_4$ and carry the same charge 
$Q'=-\frac 14$, as shown in Table~\ref{tab:ex1to3}. The third family fermions receive masses  from the  tree-level 
couplings 
\[ \lambda_t 10_t 10_t 5_{H_u}  +\lambda_b 10_t\bar 5_4 \bar 5_{H_d} \ .
\]  
We observe that there are more than one families on the same matter curves and  the present configuration 
implies a  single  intersection point for the corresponding Yukawa couplings. In addition, by turning on non-zero fluxes we obtain  
hierarchical textures for the quark and charged lepton fermion mass matrices as described in~\cite{Cecotti:2009zf}.

 The vector-like family is accommodated as follows: Left-handed doublet  and right-handed up quarks together 
 with right-handed electrons, are found in the $\overline{10}_4+10_2$ pair. Furthermore, the lepton doublet in $5_3$
  and the colour triplet in $5_6$ form a complete  fiveplet, and, together with $\bar 5_5$ constitute a pair  in 
  $\bar 5_5+(5_6+5_3)$ which  completes the vector-like family. These extra states form the following trilinear couplings 
  involving singlet fields
 \bea
\overline{10}_4 10_2\theta_{53}+\bar 5_{5}5_3\theta_{13}+\bar 5_{5}5_6\theta_{43}\label{vecmasses}
\eea

It is possible that these singlets  acquire vevs just below the GUT scale and give masses of this order to vector-like
 fields  which eventually decouple from the spectrum. Indeed, this would happen, for example, in the case of an anomalous 
 $U(1)'$ symmetry where the  vevs of the singlets  are fixed to be close to the GUT scale by virtue of the Green-Schwarz (GS)
 mechanism. In this case, the $U(1)'$ symmetry breaking scale is also associated with the GUT scale  and the $Z'$ boson
 becomes superheavy so that any New Physics effects from its non-universal couplings are highly suppressed. Yet, 
 it is possible that the GS mechanism is realised with large  vevs not involving all or some of the singlets in~(\ref{vecmasses}).  
 This allows the possibility to freely choose low scale vevs for these  latter singlet fields and obtain TeV masses to the 
vector-like fields, giving rise to interesting phenomenological implications.  However, if these vector-like 
 fields are to be relevant for the anomalies in the $B$-decays, the  singlets  involved in these terms should acquire non-zero 
  vevs $\langle \theta_{ij}\rangle$ of the order ~few TeV.  Besides, there are also couplings of  the vector-like family with 
   the three generations such as  $\overline{10}_4 10_{3}\theta_{54}+\overline{10}_4 10_t\theta_{51}$ etc. 
  Assuming this then, after mixing with the vector-like family, this model can lead to non-universal couplings of the 
  light mixed states which can be selected in such a way as to account for the recent hint for   anomalous $B$-decays,
   along the lines of~\cite{King:2017anf}.
    Therefore, we  conclude that a scenario such  as~\cite{King:2017anf} can, in principle, be realised in this fluxed GUT framework.

\section{Conclusions}

In this work, we have presented a class of  fluxed  $SU(5)$ GUTs which  naturally incorporate an abelian symmetry, $U(1)'$, 
 where the associated gauge boson $Z'$ displays non-universal gauge couplings to fermion families. 
This paper represents the first study of non-universal $Z'$ gauge bosons in F-theory phenomenology. In our analysis we have
 considered an F-theory inspired framework where  $SU(5)$ GUT emerges as a subgroup of the maximal exceptional gauge symmetry
 $E_8\supset SU(5)\times SU(5)_{\perp}$. Non-trivial universal fluxes are assumed along the Cartan generators of the custodial 
 group factor, $U(1)_{\perp}^4\subset SU(5)_{\perp}$, breaking the symmetry and generating chirality of the $SU(5)$ spectrum. 
A ${\cal Z}_2$ monodromy is imposed which allows rank-one fermion mass matrices and reduces the original  symmetry to 
  $SU(5)\times U(1)_{\perp}^3 $, while a hypercharge flux breaks the $SU(5)$ GUT down to the Standard Model and, at the same time, 
induces the observed fermion chirality to the low energy spectrum.  We then assume that the $U(1)_{\perp}^3 $ gauge group is broken 
to a single non-universal $U(1)'$, by some unspecified Higgsing mechanism.

We have analysed the case where the non-universal $U(1)'$ is
an anomaly-free  linear combination of the remaining three $U(1)_{\perp}$ factors, $U(1)'=\sum_{i=1}^3 c_i U(1)_{\perp,i}$.
We chose a basis where two of $ U(1)_{\perp,i}$ symmetries are identified with $U(1)_{\psi}\in E_6$ and $U(1)_{\chi}\in SO(10)$
and we formulated the gauge anomaly cancellation conditions in terms of a set of integers $M_a, N_b$ introduced to parametrise 
the universal  and hypercharge flux effects.  We have scanned for solutions for the three coefficients $c_i$ to determine the multiplicities
of the corresponding light spectra in terms of the flux integers $M_a, N_b$. We found a plethora of models which are classified
with respect to the nature of the non-universal couplings to fermions and the emergent spectrum. There are several classes  
of models with the minimal low energy MSSM spectrum, as well as several classes with additional vector-like fields. 

In particular we highlight
the phenomenologically promising classes with universal couplings to the three chiral fermion generations with only the 
single vector-like family having non-universal couplings. 
When the vector-like family mixes with the chiral families,
non-universality involving the mixed light states may result.
Such models naturally suppress New Physics contributions and models can be found which are universal in the the first two light quark families,
compatible with known processes such as the $K^0-\bar K^0$ mixing, while involving non-universality in the third family.
In principle these types of models may alter the branching ratios of the 
$B$-decays in accordance with the recently observed deficit in the $\mu^+\mu^-$ channel.
We discuss the plausibility of the above scenario in relation the anomaly cancellation conditions and the
scale of the New Physics effects.

\subsection*{Acknowledgements}
S.\,F.\,K. acknowledges the STFC Consolidated Grant ST/L000296/1.
This project has received funding from the European Union's Horizon 2020 research and innovation programme under the Marie Sk\l{}odowska-Curie grant agreements 
Elusives ITN No.\ 674896 and
InvisiblesPlus RISE No.\ 690575.

\appendix

\section{Singlets}

For completeness, in this short appendix  we present the
singlet fields with their `charge' assignments.
\begin{table}[H]
	\small
	\centering
	\begin{tabular}{|c|c|c|c|}
		\hline
	Singlet Field & Weights   & $Q^\prime$                                                                                    & Multiplicity  \\ \hline
	$\theta_{13}$ & $t_1-t_3$ & $\frac{\sqrt{3}c_1}{2}$                                                                    & $M_{\theta_{13}}$ \\
	$\theta_{14}$ & $t_1-t_4$ & $\frac{c_1+2 \sqrt{2} c_2}{2 \sqrt{3}}$                                        & $M_{\theta_{14}}$ \\
	$\theta_{15}$ & $t_1-t_5$ & $\frac{1}{12} \left(2 \sqrt{3} c_1+\sqrt{6} c_2+3 \sqrt{10} c_3\right)$  & $M_{\theta_{15}}$ \\
	$\theta_{34}$ & $t_3-t_4$ & $\frac{\sqrt{2} c_2-c_1}{\sqrt{3}}                                          $  & $M_{\theta_{34}}$ \\
	$\theta_{35}$ & $t_3-t_5$ & $\frac{1}{12} \left(-4 \sqrt{3} c_1+\sqrt{6} c_2+3 \sqrt{10} c_3\right)$ & $M_{\theta_{35}}$ \\
	$\theta_{45}$ & $t_4-t_5$ & $\frac{1}{4} \left(\sqrt{10} c_3-\sqrt{6} c_2\right)                        $  & $M_{\theta_{45}}$ \\
	$\theta_{31}$ & $t_3-t_1$ & $-\frac{1}{2} \left(\sqrt{3} c_1\right)                                           $  & $M_{\theta_{31}}$ \\
	$\theta_{41}$ & $t_4-t_1$ & $-\frac{c_1+2 \sqrt{2} c_2}{2 \sqrt{3}}                                     $  & $M_{\theta_{41}}$ \\
	$\theta_{51}$ & $t_5-t_1$ & $\frac{1}{12} \left(-2 \sqrt{3} c_1-\sqrt{6} c_2-3 \sqrt{10} c_3\right)$ & $M_{\theta_{51}}$ \\
	$\theta_{43}$ & $t_4-t_3$ & $\frac{c_1-\sqrt{2} c_2}{\sqrt{3}}                                          $  & $M_{\theta_{43}}$ \\
	$\theta_{53}$ & $t_5-t_3$ & $\frac{1}{12} \left(4 \sqrt{3} c_1-\sqrt{6} c_2-3 \sqrt{10} c_3\right)$  & $M_{\theta_{53}}$ \\
	$\theta_{54}$ & $t_5-t_4$ & $\frac{1}{4} \left(\sqrt{6} c_2-\sqrt{10} c_3\right)                        $  & $M_{\theta_{54}}$ \\ \hline
	\end{tabular}
	\caption{Master table for singlet fields}
	\label{tab:singlets}
	\end{table}

	The superpotential involving only the singlet fields reads 
	\bea 
	{\cal W}&=& \mu_{ij} \theta_{ij}\theta_{ji}+\lambda_{ijk}\theta_{ij}\theta_{jk}\theta_{ki}
	\label{supsing}
	\eea 
	where for each term $i\ne j\ne k\ne i$. The F-flatness conditions can be readily computed and obtain the 
	following compact form
	\[\frac{\partial \cal W}{\partial \theta_{ij}}= \mu_{ij} \theta_{ji}+\lambda_{ijk}\theta_{jk}\theta_{ki}
	\]

\end{document}